\documentclass[12pt,a4paper]{article}

\usepackage[english]{babel}
\usepackage{standalone}
\usepackage{float}
\usepackage{booktabs}
\usepackage{amsmath}
\usepackage{rotating}
\usepackage{graphics}
\usepackage{listings}
\usepackage{setspace}
\usepackage[square,numbers,sort&compress]{natbib}
\bibpunct{[}{]}{,}{a}{,}{,}
\setcitestyle{numbers}
\usepackage[titletoc]{appendix}
\usepackage{cite}
\usepackage[left=1.5cm,right=1.5cm,top=2cm,bottom=2cm]{geometry}
\usepackage{url}
\usepackage[titletoc]{appendix}
\graphicspath{{image/}}
\usepackage{tikz}
\usetikzlibrary{fit,positioning}
\usetikzlibrary{bayesnet}
\usetikzlibrary{shapes}
\usetikzlibrary{arrows, decorations.markings, matrix}
\usetikzlibrary{fit}
\usetikzlibrary{chains}
\usetikzlibrary{arrows}
\usepackage[font=footnotesize, labelfont={bf}, margin=1cm]{caption}
\usetikzlibrary {trees}
\usetikzlibrary{arrows,positioning,shapes.geometric}
\usepackage{forest}
\usepackage{tikz-qtree}
\usepackage{tabularx}
\usepackage{array}
\usepackage{colortbl}
\usepackage{enumerate,multicol}
\usepackage{amsmath, amssymb, amsthm}
\usepackage{enumitem,bm}
\usepackage[para,online,flushleft]{threeparttable}
\usepackage{parcolumns}

\usepackage[multiple]{footmisc}
\usepackage[sectionbib]{chapterbib}
\usepackage{etoolbox}
\usepackage{authblk}
\usepackage[position=bottom]{subfig}
\usepackage{soul,color}
\usepackage{pdfpages}
\usepackage{multirow}
\usepackage{makecell}
\usepackage{fancyvrb} 
\usepackage{moreverb}

\newcommand{\textcite}[1]{\citeauthor{#1}~\citeyear{#1}}

\begin{document}

\title{A review of heath economic evaluation practice in the Netherlands: are we moving forward?}
\author[1]{Andrea Gabrio\thanks{E-mail: a.gabrio@maastrichtuniversity.nl}}
\affil[1]{\small \textit{Department of Methodology and Statistics, Faculty of Health Medicine and Life Science, Maastricht University, P. Debyeplein 1, 6229 HA Maastricht, NL.}}

\date{}
\maketitle
\hrule

\newpage 
\abstract{In 2016, the Dutch National Health Care Institute issued new guidelines that aggregated and updated previous recommendations on key elements for conducting economic evaluation. However, the impact on standard practice after the introduction of the guidelines in terms of design, methodology and reporting choices, is still uncertain. To assess this impact, we examine and compare key analysis components of economic evaluations conducted in the Netherlands before (2010-2015) and after (2016-2020) the introduction of the guidelines. We specifically focus on two aspects of the analysis that are crucial in determining the plausibility of the results: statistical methodology and missing data handling. Our review shows how many components of economic evaluations have changed in accordance with the new recommendations towards more transparent and advanced analytic approaches. However, potential limitations are identified in terms of the statistical software and information provided to support the choice of missing data~methods.}

\begin{description}
\item[Keywords.] economic evaluations; review; cost-effectiveness; analytic approaches; The Netherlands
\item[Classification codes.] D61; D70; D81; H51; I18
\item[Funding.] This research did not receive any specific grant from funding agencies in the public, commercial, or
not-for-profit sectors.
\item[Conflict of interest.] The author(s) declared no potential conflicts of interest with respect to the research, authorship, and/or publication of this article.
\end{description}



\clearpage

\section{Introduction}\label{intro}
Health economic evaluation is a relatively new discipline whose definition and application have gradually but constantly evolved during the last decades. Nowadays, economic evaluations are primarily conducted to inform decisions about the allocation of limited resources across a pool of alternative healthcare interventions within a given health care system. The first official adoption of economic evaluation within a national public healthcare system is attributed to the Australian government~\citep{PBAC1992} in the early '90s, and later followed by other public authorities in many other countries~\citep{hjelmgren2001health}. Although the purpose of economic evaluation remains the same across different jurisdictions, the presence of geographical and socio-cultural differences imposes national pharmaceutical decision-making committees to define their own requirements and guidelines for pharmacoeconomic evaluations~\citep{ISPOR}. 

In the Netherlands, the Dutch National Health Care Institute (\textit{Zorginstituut Nederland} or ZIN) is the body in charge of issuing recommendations and guidance on good practice in economic evaluation, not just for pharmaceutical products, but also in relation to other fields of application that include prevention, diagnostics, medical devices, long-term care and forensics. In 2016, ZIN issued an update on the guidance for economic evaluation~\citep{nederland2016guideline}, which aggregated into a single document and revised three separately published guidelines for pharmacoeconomic evaluation~\citep{postma2006farmaco}, outcomes research~\citep{delwel2008guidance} and costing manual~\citep{hakkaart2015methodologie}. The novel aspects and future policy direction introduced by these guidelines have already been object of discussion, particularly with respect to the potential impact and concerns associated with their implementation in standard health economic practice in the Netherlands~\citep{garattini2017dutch, versteegh2016good}. Given the importance covered by these guidelines, an assessment  of their impact on economic evaluation practice in the Netherlands would allow to draw some conclusions.

Our objective was to review the evolution of economic evaluation practice in the Netherlands before and after the introduction of the ZIN's 2016 guidelines. In addition, we provide an in-depth assessment of the quantitative approaches used by analysts with a focus on the statistical methods, missing data methods and software implemented. Given the intrinsic complexity that characterise the analysis framework of health economic data, the choice of the analytical approaches to deal with these problems as well as transparent information on their implementation is crucial in determining the degree of confidence that decision-makers should have towards cost-effectiveness results obtained from such analyses~\citep{ramsey2015cost}.

The rest of the article is structured as follows. Section~\ref{guidance} briefly outlines the key elements of the ZIN's 2016 guidelines, with a focus on the changes that were introduced with respect to previous guidance. Section~\ref{review} presents the review methodology and compares the characteristics of the studies with the recommendations from the 2016 guidelines. Section~\ref{analysis} reviews the analytical methods and software used, while Section~\ref{missing} focuses on the choice of missing data methods and uses a structured grading scheme to evaluate the studies based on the overall level of missingness information provided. Finally, Section~\ref{conclusions} summarises our findings and recommendations for future research.

\section{The ZIN 2016 guidelines}\label{guidance}
The main objective of the guidelines is to ensure the comparability and quality of economic evaluations in the Netherlands, therefore facilitating the task of the decision-maker regarding the funding or reimbursement of new healthcare interventions. Following the example of guidelines issued by decision-making bodies in other countries, including the National Institute for Health and Care Excellence in the UK~\citep{excellence2013guide}, the recommended features for economic evaluations are summarised in a "reference case", although deviations from it are allowed when properly justified (e.g.~in case of non-pharmaceutical products). 

Based on the structure of the reference case, four essential topics are briefly summarised: framework, analytic approach, input data and reporting. We do not review information related to cost-benefit, cost-minimisation or budget impact analyses as these do not fall within the scope of this article. For a thorough examination of the guidelines and implication on practice we refer the interested reader to, respectively, the original document~\citep{nederland2016guideline} and two recent articles~\citep{garattini2017dutch, versteegh2016good}. 

\subsection{Framework of the economic evaluation}\label{framework}
A series of elements form the framework and allow to identify the objective and the \textit{users} of the economic evaluation. According to the reference case, the mandatory \textit{perspective} to be adopted is the societal perspective, which implies that all costs and benefits, irrespective of who is the bearer/beneficiary, should be taken into account. Results from other perspectives (e.g.~healthcare provider) may also be presented as additional analyses. The \textit{research question} is summarised by the PICOT (Patient, Intervention, Control, Outcome and Time) criteria and should involve: a population in the Dutch setting (P); a new healthcare intervention (I) and standard of care (C) that can be applied in the Netherlands; pre-defined outcome measures (e.g. clinical, patient-reported); the expected lifetime of the target population (T). It is also recommended to "scope" the PICOT criteria beforehand with the relevant stakeholders (e.g. patient organisations) to benefit from their expertise and experience~\citep{nederland2015beoordeling}.

\subsection{Analytic approach}\label{analytic}
The number and type of analytic techniques that should be implemented depend on the type of economic evaluation. \textit{Cost-Effectiveness Analysis} (CEA) and \textit{Cost-Utility Analysis} (CUA), respectively based on clinical or Quality-Adjusted Life Years (QALYs) measures, are the most popular types of analyses, with CUA being the preferred choice since it allows better comparability of results between different health conditions. 

\textit{Discounting} should always be applied when outcome data are analysed over a time horizon exceeding one year using a yearly discount rate of $1.5\%$ for effects and $4\%$ for costs. \textit{Uncertainty} surrounding the economic results from the analysis should always be assessed to: 1) quantify the impact on cost-effectiveness conclusions; 2) determine if and how much additional research may reduce uncertainty. The methods and type of uncertainty analyses vary according to the type of economic evaluation, with a clear distinction between empirical (e.g.~CUA alongside a trial) and model-based (e.g.~simulation models) analyses since the type of input data and objective are~different.

Empirical analyses should implement statistical methods to quantify the uncertainty around mean incremental costs, effects and cost-effectiveness ratios (ICERs). Bootstrapping is the standard approach used to generate a large number of resampling draws and quantify the uncertainty through the computation of confidence intervals, Cost-Effectiveness Planes~\citep[CEP;][]{black1990plane} and  Cost-Effectiveness Acceptability Curves~\citep[CEAC;][]{van1994costs}. Appropriate statistical methods should also be used to quantify the impact of \textit{missing data} uncertainty on the results, with Multiple Imputation~\citep[MI;][]{van2018flexible} being the recommended approach, with Expectation-Maximisation as a possible alternative~\citep[EM;][]{dempster1977maximum}. Regression techniques may also be used to increase the precision and correct for differences between groups, while alternative approaches can be used to assess the robustness of the results in scenario and sensitivity analyses. 

Model-based analyses often consist in patient-level simulation methods which should perform \textit{Probabilistic Sensitivity Analysis}~\citep[PSA;][]{claxton2005probabilistic} by varying the assumed distributions and associated measures of variability to assess the impact of parameter uncertainty on ICERs. In addition, deterministic sensitivity analysis should be carried out on other model inputs (e.g.~discount rate, cost prices) and structural uncertainty should be made transparent by presenting a clear overview of the model assumptions. \textit{Value of Information} (VOI) analysis should be performed and an estimate of the Expected Value of Perfect Information (EVPI) should be produced, quantifying all consequences of the uncertainty around the model parameters~\citep{claxton2006using}. Model validation is crucial and should provide information on the model structure, input data and software code.

\subsection{Input data}\label{input}
In empirical analyses, input data on \textit{clinical effectiveness} are collected and derived from the study, whereas in model-based analyses the clinical effectiveness data need to be underpinned by a systematic review of the literature, preferably using evidence from randomised studies and head-to-head comparisons. Identification, measurement and valuation of \textit{cost data} should be done following the guidance in the costing manual~\citep{hakkaart2015methodologie}. All relevant societal cost should be identified, including those related to the healthcare system (direct and indirect medical costs), patient and family (e.g.~travel, informal care), other sectors (e.g.~volunteering) and productivity losses (e.g.~due to absenteeism). The \textit{friction method} should be used to calculate productivity losses as a result of paid work absenteeism. Costs are computed by multiplying the volume of a specific service (i.e.~resource use information collected during the trial) with the corresponding standardised national unit price and adjusting for inflation via consumer price index. \textit{Quality of life} data should be collected by means of validated, generic quality-of-life self-reported questionnaires which assign to each patient a utility score valued based on the preferences of the general population in the country. The reference case identifies the EQ-5D-5L questionnaire~\citep{janssen2013measurement} as the preferred instrument to measure quality of life, valued through Dutch reference values~\citep{versteegh2016dutch}. Alternative questionnaires and other methods to evaluate quality of life may be added next to the reference case.

\subsection{Reporting}\label{reporting}
Information related to input data (effectiveness, costs and quality of life) should be reported in a transparent way. This includes, but is not limited to, details of studies used to retrieve effectiveness data (e.g.~patient characteristics), prices and volumes of all cost components, questionnaires or valuation methods for quality of life data. For economic evaluations based on empirical studies, \textit{missing data information} should be clearly reported in terms of amount, whether partially-observed individuals differed from completers and whether missingness was addressed at the study design. Alternative approaches should be implemented to assess the sensitivity of the results to different methods. 

The reporting of the results should be tailored to the type of analysis performed, namely either base-case or uncertainty analysis. In the \textit{base-case analysis} both the total and incremental costs/effects for each intervention group should be reported alongside the ICER, and graphically represented via the CEP. In \textit{uncertainty analysis}, parameter uncertainty should be reported in terms of the minimum and maximum variations of the ICER, as well as the impact on the incremental costs and effects via tabular form and graphically by means of a tornado diagram. Results of PSA (model-based) or bootstrapping (empirical) should be presented graphically via CEP and CEAC. As an alternative, results under the net benefit approach~\citep{stinnett1998net} for each intervention can be reported. Finally, results of VOI analysis should be presented using different reference values of the ICER.

\section{Literature Review}\label{review}
We identified papers within the period 1 Jan 2010 to 31 December 2020. Articles were considered eligible for the review only if they were cost-effectiveness or cost-utility analyses targeting a Dutch population. Study protocols, pilot studies as well as cost-benefit, cost-minimisation or budject impact analyses were excluded. We relied on the search engines of two online full-text journal repositories: 1) \texttt{pubmed}, 2) \texttt{Zorginstituut}. The key words used in the search strategy were (\texttt{cost-effectiveness} OR \texttt{cost-utility} OR \texttt{economic evaluation}). The on-line databases identified $4319$ articles most of which were duplicates. After abstract review, $647$ articles were considered, of which $190$ fulfilled the eligibility criteria. We report the full list of reviewed studies in the online Appenidx 

\subsection{Review}\label{descriptives}
We present and compare the articles reviewed between two separate periods (2010-2015 and 2016-2020) to assess and identify changes in standard health economic practice after the introduction of the ZIN's 2016 guidelines. We summarised key results in terms of the type of analysis and analytic approaches implemented. With regard to empirical analyses, we looked in detail at the statistical methods and software used, while also reviewing and evaluating the strategies implemented to handle missing data.

Table~\ref{tab1} reports information about the reviewed studies, separately between the 2010-2015 and 2016-2020 periods, and compares it to the recommendations on each element of the economic evaluation as described in the reference case of the 2016 guidelines.
\begin{center}
TABLE 1 
\end{center}\vspace{0.5cm}

Out of the $190$ studies included, about half were published between 2010-2015 ($96$) and between 2016-2020 ($94$), with also comparable numbers in terms of empirical ($86$ vs $80$) as well as model-based analyses ($10$ vs $14$). In the Appendix, we report a visual representation of the sample size distribution based on the $166$ empirical studies included in the review. 

Some considerable changes are observed between the two periods in regard to different analysis components: 1) a sensible increase in the proportion of studies adopting a societal perspective in the primary analysis and a healthcare perspective in secondary analyses (from 23\% to 40\%); 2) an increase in the proportion of studies performing CUAs as primary analysis (from 31\% to 44\%) and a decrease in the number of primary CEAs (from 30\% to 17\%); 3) an uptake in the number of studies including all relevant societal costs in the analysis (from 53\% to 68\%); 3) an increase in the proportion of CUAs which provide clear information on the EQ-5D questionnaires, for both 5L (from 4\% to 12\%) and 3L (from 24\% to 33\%) versions. 

In addition, we observe an increase in the proportion of studies following the recent guidelines in regard to the choice of the discount rates for future effects and costs (from 58\% to 68\%) as well as the use of the friction method to calculate productivity losses (from 39\% to 59\%). Limited variations are observed in the number of studies using both CEP and/or CEAC to report the results from uncertainty analysis, and the time horizon chosen in empirical analyses. Although there is a slight decrease in the proportions of model-based analyses using a lifetime horizon, these are calculated from relatively small numbers ($10$ studies between 2010-2015 and $14$ between 2016-2021) and may therefore be misleading. Finally, we observe that only one study within each period conducted VOI analysis and provided an estimate of EVPI.

\section{Analytic approaches}\label{analysis}
In this section we explore in more detail the information provided by the reviewed studies in relation to type of analytical approaches used to perform the economic evaluation and assess uncertainty. We also review information concerning the specific software program used as it may provide insights on practitioners' preferences of implementation and potential room of improvement. We specifically focus on the choice of the statistical approaches as it represents a crucial element in any economic evaluation to determine the validity and reliability of cost-effectiveness conclusions.

\subsection{Statistical methods}\label{stats}
We begin by reviewing the type of statistical methods used to estimate mean incremental costs and effectiveness between treatment groups (and ICERs) and to quantify the level of uncertainty around the estimated quantities. According to ZIN's 2016 guidelines and current literature, for empirical analyses, bootstrapping is the recommended approach to deal with non-normal distributions and quantify the level of uncertainty around the incremental mean cost and effect estimates~\citep{campbell1999bootstrapping}. Regression technique are also important in order to obtain adjusted estimates and to control for potential imbalances in some baseline variables between treatment groups~\citep{manca2005estimating, van2009deal}. 

\begin{figure}[!h]
\centering
\scalebox{0.6}{
\includegraphics{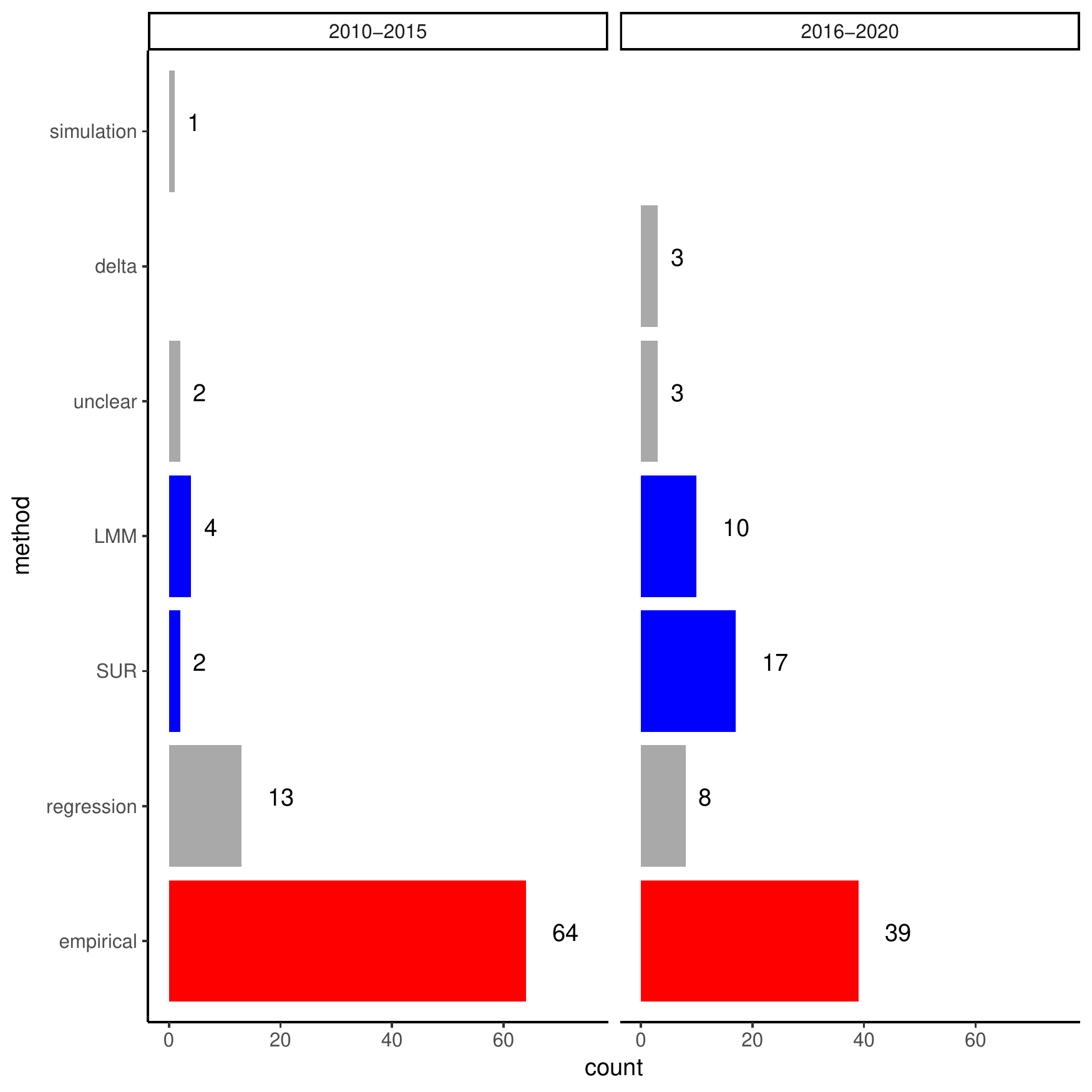}
}
\caption{Barchart of the number of empirical studies grouped by statistical methods implemented. Results are distinguished by time period (2010-2015 and 2016-2020) and grouped using the following method's classes: simulation, delta method, unclear, linear mixed effects model (LMM), seemingly unrelated regression (SUR), regression, empirical.}\label{fig1}
\end{figure}

Almost all reviewed empirical studies used bootstrapping (95\%) although with different choices for the number of iterations: the mean and standard deviation of the bootstrap replications, computed over the studies which provided such information (86\%), were $4321$ and $5883$, respectively, with the most popular choices being $5000$ (55\%) followed by $2000$ (29\%). Studies showed even more variability in the methods used in combination with bootstrapping to correct for potential sources of bias. Figure~\ref{fig1} shows the type of statistical techniques implemented among the $166$ empirical analyses in our review.

\begin{center}
FIGURE 1 
\end{center}\vspace{0.5cm}

Seven general classes of statistical approaches were identified, among which the empirical analysis without any adjustment was the most popular choice across both time periods. Regression-based adjustment methods were also widely used either in the form of: simple univariate regression adjustment~\citep{manca2005estimating}; bivariate regression adjustment accounting for the correlation between effects and costs, also known as \textit{seemingly unrelated regression}~\citep[SUR;][]{zellner1962further}; linear mixed modelling to account for clustering effects, e.g.~in cluster randomised trials~\citep{rice1997multilevel, manca2005assessing}. Finally, delta adjustment~\citep{vickers2001analysing, van2009deal} or simulation methods were only rarely adopted. It is apparent how between the two periods there was a shift in the use of the methods, with a considerable decrease of about 40\% in the number of analyses not performing any adjustment (red bars), in contrast to an uptake in the number of analyses using SURs (from $2$ to $17$) or LMMs (from $4$ to $10$) adjustments (blue bars). Although these methods are not explicitly mentioned in the 2016 guidelines, the need to perform regression adjustment was clearly indicated as an important component in empirical analyses and both LMMs and SURs are widely used methods among the international health economics literature~\citep{willan2004regression}. Bootstrapped confidence intervals for the estimated mean incremental outcomes were calculated for all analyses, although only $53$ studies (32\%) provided information on the specific methods used. Among those providing such information, $29$ (55\%) applied bias-adjusted and accelerated methods~\citep{efron1994introduction} and $24$ (45\%) applied standard percentile methods.

For model-based analyses, \textit{Monte Carlo} methods~\citep{briggs1999handling} are the standard algorithms used in decision analytic models to simulate the evolution/progression of a target patient population and to aggregate over time the total quality of life and costs associated with each patient profile (e.g.~via multi-state or Markov models). Among the $190$ reviewed studies, only $24$ were model-based analyses (see Table~\ref{tab1}) and all used Monte Carlo simulation methods. The vast majority of the approaches were Markov models (88\%), followed by a decision tree and some unclear specifications, with no considerable differences between the two time periods. Information on model implementation was provided by about 75\% of the studies, with a mean number of iterations run of $3028$, standard deviation of $3089$, and with the most popular choice being $1000$. For Markov models, the number of assumed health states varied from $2$ to $12$, with cycle lengths ranging from $1$ up to $12$ months, although considerable variability was observed across the analyses.

\subsection{Software}\label{software}
\begin{figure}[!h]
\centering
\scalebox{0.6}{
\includegraphics{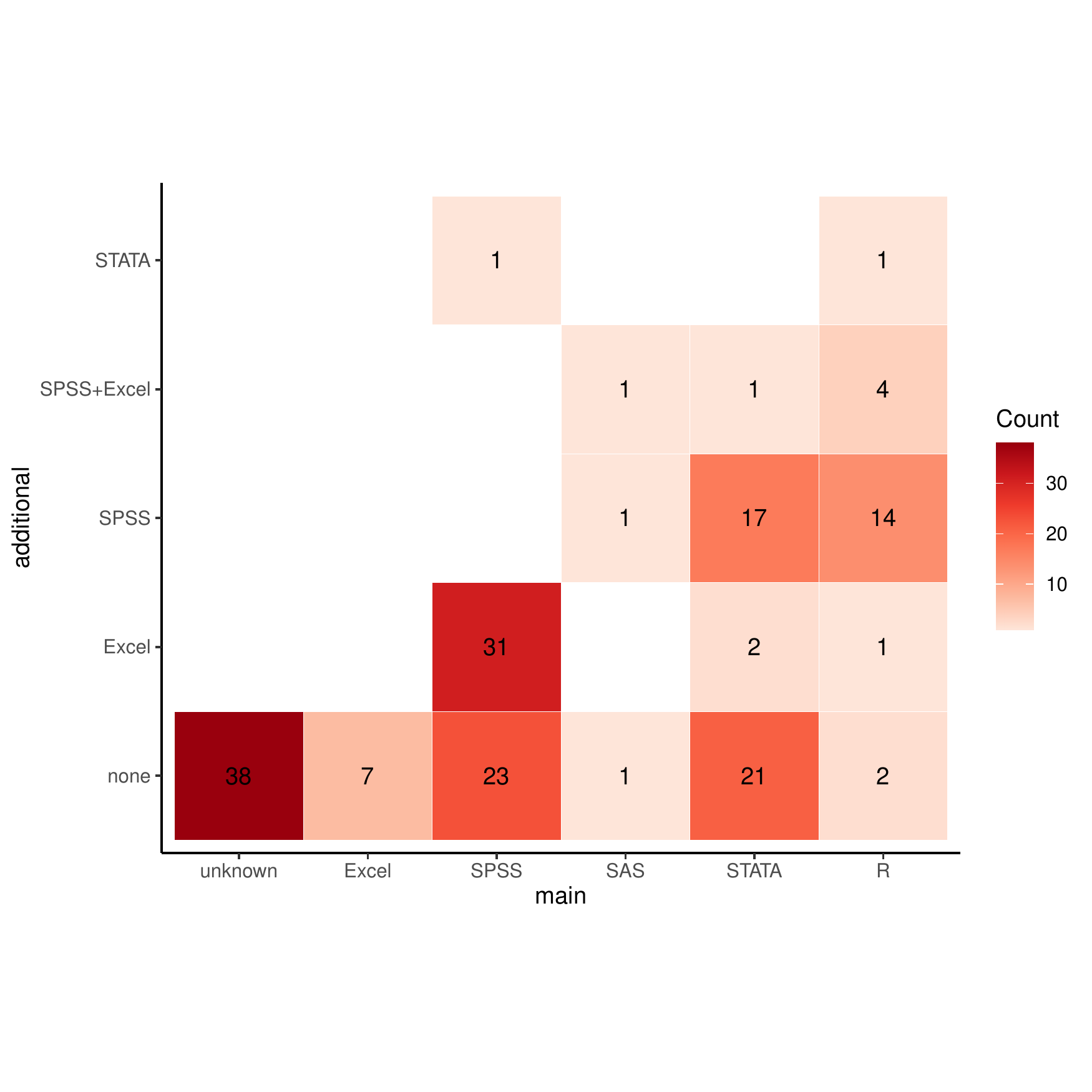}
}
\caption{Heatmap of the type of software programs used among the empirical analyses. Software use is distinguished between main and additional analyses, defined according to the associated order or tasks that was specified in the information obtained from the studies. For each pairwise (main-additional) software combination, darker-coloured squares are associated with higher frequencies of use compared to lighter-coloured squares.}\label{fig2}
\end{figure}

We looked at the different type and combination of software programs used as an indication of the implementation preferences of analysts when performing economic evaluations. Since no considerable differences were observed when comparing software use over time, we present the results across all publication years from 2010 up to 2020, but divided by type of analysis (empirical and model-based). Figure~\ref{fig2} shows an heatmap of the type of software used among the $166$ empirical studies included in the review. Software programs are distinguished into "main" and "additional" categories according to the order (e.g.~first mentioned) or tasks (e.g.~main analysis vs secondary analyses) for which they were used according to the information provided by each study. 

\begin{center}
FIGURE 2
\end{center}\vspace{0.5cm}

The most popular software was \texttt{SPSS}, chosen by $87$ (52\%) of the studies, either in the main (33\%) or additional (19\%) analysis, and often used in combination with \texttt{Excel} or by itself. When either \texttt{STATA} (26\%) or \texttt{R} (13\%) were used in the main analysis, \texttt{SPSS} was still the most popular choice in additional analyses. Other combinations of software were less frequently chosen, even though $38$ (23\%) of the studies were unclear about the software implemented.

Among the $24$ model-based analyses, $14$ (58\%) did not provide any information in regard to the choice of the software, while \texttt{Excel} alone was the most frequent software choice in $9$ (38\%) studies, followed by \texttt{TreeAge} with $2$ (8\%), and \texttt{R}, \texttt{Delphi} and \texttt{SPSS} with $1$ (all $<5\%$).  

\section{Missing data methods}\label{missing}
The choice of the statistical methods to handle missing data has a potentially large impact on cost-effectiveness results and should be made to avoid implausible assumptions, which may lead to incorrect inferences. Since it is never possible to check assumptions about unobserved values, unless the amount of missing data is negligible (e.g.~$<5\%$), a principled approach is typically recommended. This amounts to perform the analysis using a method associated to a benchmark missing data assumption (base-case analysis), and then assess the robustness of the base-case results to alternative assumptions using different methods (sensitivity analysis). It is important that both base-case and sensitivity analyses implement methods that are based on "plausible" missingness assumptions to ensure that the impact of missing data uncertainty is adequately quantified~\citep{molenberghs2007missing}.

By their own nature missing data represent a crucial problem in empirical analyses but are less relevant in the context of model-based analyses. Within the second class of models, the long-term extrapolation of outcome data (e.g.~survival beyond observed time horizon) represents a similar problem and is often accomplished through parametric or non-parametric methods. However, for the purpose of this review, we will exclusively focus on standard missing data methodology implemented in empirical analyses which represents the majority of the reviewed economic evaluations. 

\subsection{Base-case and sensitivity analysis}\label{missing_methods}
We first review the type of missing data methods implemented among empirical analyses. These are also distinguished by time period and by whether they were implemented in the base-case analysis (method used in the main analysis) or in sensitivity analysis (alternative methods used to check the robustness of base-case results). We initially planned to report missing data information separately by effects and costs but, after reviewing the analyses, we noted that only a small number of studies provided this level of detail. In the following, we will therefore provide results under the assumption that the same approaches were used to handle both missing effects and~costs. In the Appendix, we provide a visual representation of the distribution of missing effect and cost rates based on the empirical studies which provided this information.

\begin{figure}[!h]
\centering
\scalebox{0.6}{
\includegraphics{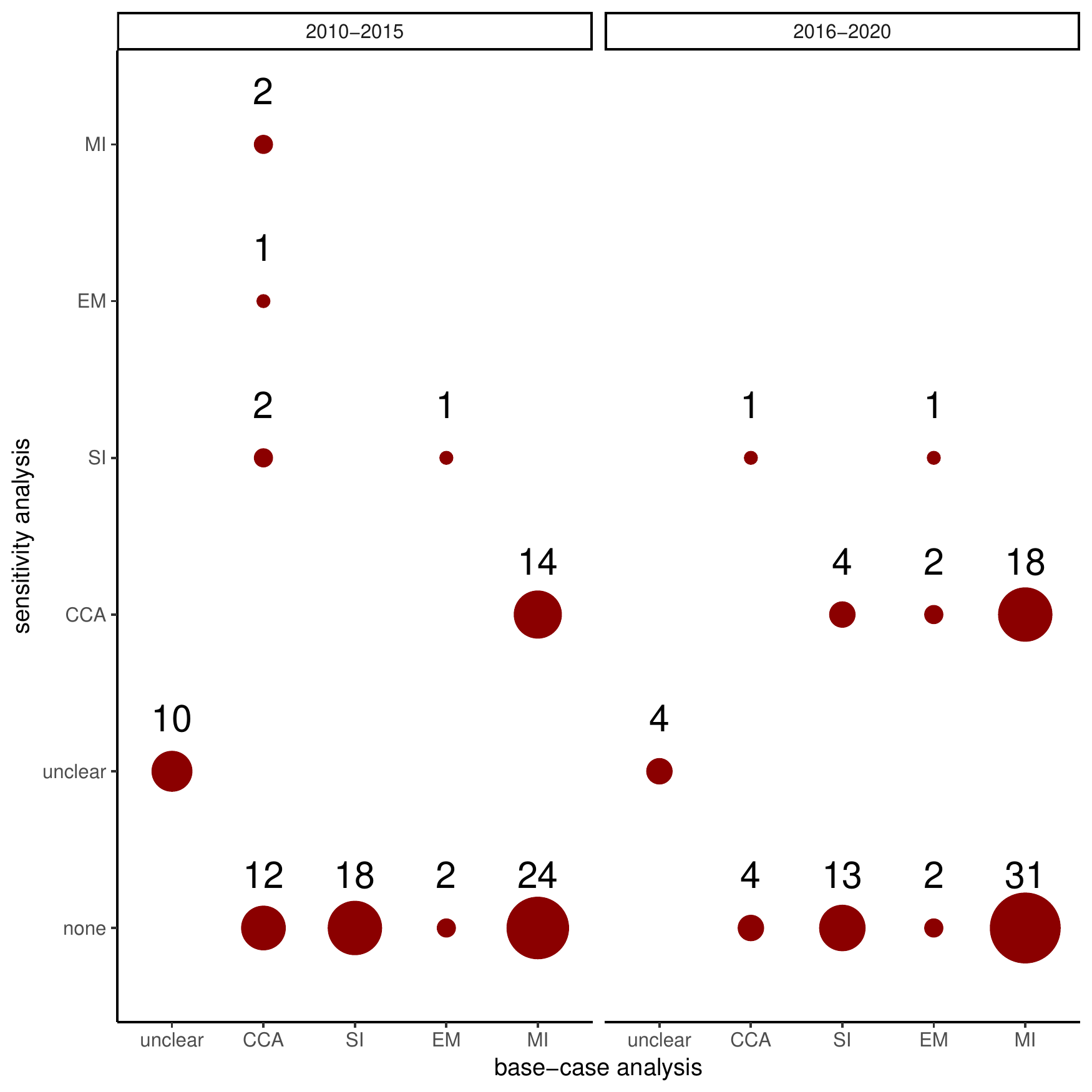}
}
\caption{Bubble plot of the type of missing data methods used among the empirical studies. Results are distinguished by time period (2010-2015 and 2016-2020) and grouped into $6$ categories: no method (none); unclear (unclear); complete case analysis (CCA); single imputation (SI); expectation-maximisation (EM); multiple imputation (MI). Methods are distinguished according to whether they were used in the base-case or sensitivity analysis with the size of each bubble representing the frequency of use for each combination of base-case and sensitivity analysis missing data method.}\label{fig3}
\end{figure}

Figure~\ref{fig3} shows, for both periods, a bubble plot for each combination of missing data methods implemented in the base-case and sensitivity analysis for empirical analyses, where the size of the bubbles indicates the frequency of use for each pairwise combination.

\begin{center}
FIGURE 3
\end{center}\vspace{0.5cm}

Overall, between the two periods, no drastic changes is observed in terms of the preference for missing data methods, with MI being the most popular base-case analysis choice, followed by complete case analysis (CCA), which remains the most popular sensitivity analysis choice. However, some changes are observed in the frequency of adoption of these methods. On the one side, the proportion of studies using MI in the base-case analysis has increased over time (from 28\% in 2010-2015 to 39\% in 2016-2020). On the other side, the proportion of studies has decreased for both CCA (from 14\% in 2020-2015 to 5\% in 2016-2020) and single imputation (SI) methods (from 21\% in 2010-2015 to 16\% in 2016-2020). The number of studies not clearly reporting the methods used to handle missing data has also decreased (from 12\% in 2010-2015 to 5\% in 2016-2020), while the use of other methods has not varied~notably. 

The observed trend between the two periods may be the result of the specific recommendations from the 2016 guidelines in regard to the "optimal" missing data strategy, resulting in a more frequent adoption of MI techniques and, at the same time, a less frequent use of CCA in the base-case analysis. However, in contrast to these guidelines, a large number of studies still does not perform any sensitivity analysis to missing data assumptions (about $65\%$ in 2010-2015 and $63\%$ in 2016-2020).

Information was also collected across both periods about details of MI implementation when these were provided. In particular, among the $89$ studies using MI: $50$ (56\%) used the fully-conditional or chained equation version~\citep{van2018flexible}, while the rest of the studies did not specify the version used; $32$ (36\%) used predictive mean matching as imputation technique, $10$ (11\%) used linear or logistic regression, $1$ (1\%) used predictive score matching, while the rest of the studies provided unclear information. Finally, the mean and standard deviation of the number of imputed dataset generated were $18$ and $19$, respectively, with the most frequent choice being $5$ (23\%).

\subsection{Quality of missing data information}\label{quality_scores}
We finally review the quality of the overall missing data information reported by the studies. We specifically rely on the \textit{Quality Evaluation Scheme} (QES), a structured reporting and analysis system that embeds key guidelines for missing data handling in economic evaluation~\citep{gabrio2017handling}. Detailed information about the rationale and structure of the scheme are provided in~\citet{gabrio2017handling}, while here we only provide a concise explanation for clarity.

First, a numeric score is created to reflect the amount and type of information provided on three components characterising the missing data problem: description (e.g.~number and pattern of missing data), method (e.g.~type of method and detail of implementation) and limitations (e.g.~limitations of assumptions). Each component is assigned a score weight (using a ratio $3:2:1$) according to its importance, and then summed up to obtain an overall score for each study, ranging from $0$ (no information) to $12$ (full information). Next, grades are created by grouping the scores into ordered categories from A (highest score) to E (lowest scores). Finally, studies are also grouped by type of missingness method into five ordered classes, reflecting the strength of the underneath assumptions: unknown (UNK); single imputation (SI); complete case analysis (CCA); multiple imputation/expectation maximisation (MI/EM); sensitivity analysis (SA). We note that SA represents the less restrictive method as it requires studies to justify the assumptions explored in both base-case and sensitivity analysis based on the available information. 

\begin{figure}[!h]
\centering
\scalebox{0.6}{
\includegraphics{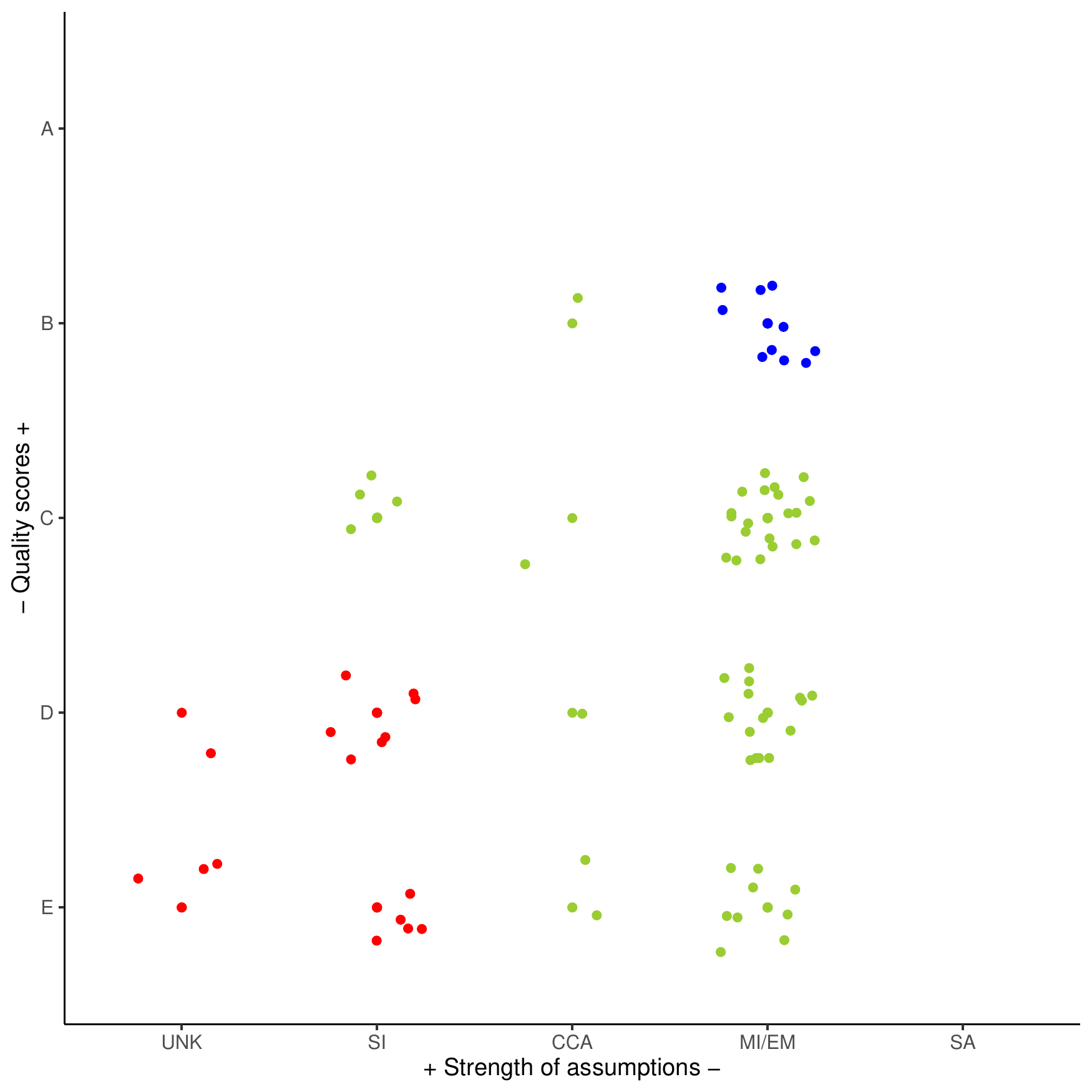}
}
\caption{Jitter scatterplot for the joint assessment of the quality of missing data assumptions and information provided by empirical studies in the period 2016-2020. The strength of missing data assumptions is represented in terms of the type of methods used to handle missing values: unknown (UNK), single imputation (SI), complete case analysis (CCA), expectation-maximisation or multiple imputation (MI), sensitivity analysis (SA). The quality of the missing data information to support the method's assumptions is measured using the scores based on the QES and graded into the ordered categories: E, D, C, B, A.}\label{fig4}
\end{figure}

Figure~\ref{fig4} shows a graphical representation of the quality scores (expressed in grades) in combination with the strength of assumptions (expressed by type of method) for each of the $80$ empirical studies in the period 2016-2020. We specifically focus on studies in the later period as we want to assess current missing data practice (after the introduction of the 2016 guidelines). 
 
\begin{center}
FIGURE 4
\end{center}\vspace{0.5cm}

Most of the studies lie in the middle and lower part of the plot, and are associated with a limited (grades D and E) or sufficient (grade C) quality of information. However, only a few of these studies rely on very strong and unjustified missing data assumptions (red dots in the bottom-down part), while the majority provides either adequate justifications or uses methods associated with weak assumptions (green dots in the middle part). Only $11$ (14\%) studies are associated with both high quality scores and less restrictive missingness assumptions (blue dots in the top-right part). No study was associated with either full information (grade A) or adequate justifications for the assumptions explored in base-case and sensitivity analysis (SA).

\section{Discussion}\label{conclusions}
The objective of this paper was to review and compare the practice of conducting economic evaluation in the Netherlands before and after the introduction of the ZIN's 2016 guidelines. We focussed on the type of analytic approaches and software used to conduct the analysis, while also examining the missing data methods used and critically appraise the studies based on the overall information provided on missingness.

\subsection{Descriptive review}\label{descriptive}
Descriptive information extracted from the reviewed studies (Table~\ref{tab1}) highlights some interesting discussion points when comparing economic evaluation practice between 2010-2015 and 2016-2020. First, most of the studies in the later period are CUA and use a societal perspective, with CEAs and alternative perspectives provided in secondary analyses. Second, studies tend to use EQ-5D instruments to measure quality of life and include all relevant types of societal costs, including productivity losses for which the friction approach is the current reference method of calculation. Finally, reporting of cost-effectiveness results often takes into account both uncertainty and probabilistic sensitivity analysis by providing either or both CEP and CEAC. 

Most of these changes are in accordance with the 2016 guidelines, which are likely to have played a role in guiding analysts and practitioners towards a clearer and more standardised way to report health economic results. However, for some components of the analysis, such a VOI analysis or time horizon, adherence to the new guidelines seems slow (although the limited number of model-based studies makes it difficult to reach clear conclusions).

\subsection{Health economic analysis}\label{analytic_approach}
The most popular methods to quantify uncertainty around cost and effect estimates are by far bootstrapping (empirical analyses) and Monte Carlo simulation (model-based analyses). However, between the two periods a shift towards the use of statistical methods to control for potential sources of bias between treatment groups was observed, with a considerable uptake in the use of SURs and LMMs in the context of empirical analyses (Figure~\ref{fig1}). These techniques are important in order to adjust for differences in baseline variables, handle clustered data, and formally take into account the correlation between costs and effects~\citep{willan2004regression}. In addition, when further issues occur (e.g.~presence of spikes in the observed data distributions), analysts should also consider the use of tailored approaches~\citep{basu2012regression, baio2014bayesian}.

The complexity of the statistical framework for health economic evaluation requires the implementation of methods that can simultaneously handle multiple statistical issues to avoid biased results and misleading cost-effectiveness conclusions. However, it is equally important that the level of complexity of the analysis model is reflected in the way uncertainty surrounding the estimates is generated. For example, if clustered data are handled by means of LMMs, then clustered bootstrap methods should be used to properly generate resampling draws. Among all reviewed studies, we identified $13$ cluster randomised trials but only $10$ took into account clustering at the analysis stage and, among these, only $1$ study implemented clustered bootstrap methods. 

We believe that these inconsistencies are due to either limited familiarity of practitioners with advanced statistical methods or potential limitations of the software used to conduct the analysis. This seems to be supported by the fact that a considerable amount of studies still rely on non-statistical software (e.g.~\texttt{Excel}), or a combination of these and user-friendly statistical software (e.g.~\texttt{SPSS}) to perform the analysis (Figure~\ref{fig2}). Although this does not represent an issue per se, it may become problematic and potentially lead to difficult-to-spot errors when performing complex analyses without the use of more advanced and flexible software programs, such as \texttt{R} or \texttt{STATA}~\citep{incerti2019r}. 

\subsection{Missing data}\label{missing_approach}
Multiple imputation is the default method of choice for handling missing data in economic evaluations. The transition between 2010-2015 and 2016-2020 suggests an increase in the use of MI techniques in the base-case analysis together with a decrease in the use of CCA (Figure~\ref{fig3}). This suggests how analysts have become aware of the inherent limitations and potential bias of CCA and shifted towards MI as reference method. Nevertheless, improvements in the approach to deal with missing data are still needed given that many studies (more than 60\%) performed the analysis under a single missing data assumption.

This is not ideal since by definition missing data assumptions can never be checked, making the results obtained under a specific method (i.e.~assumption) potentially biased. For example, MI is often implemented under a \textit{Missing At Random} or MAR assumption (i.e.~missingness only depends on observed data). However, there is no way to test if MAR is appropriate and it is always possible that missingness depends on some unobserved quantities, corresponding to a so-called \textit{Missing Not At Random} or MNAR assumption~\citep{rubin2004multiple}. 

This is why sensitivity analysis has a crucial role in assessing the robustness of the results to a range of plausible departures from the benchmark assumption chosen in the base-case, including MNAR~\citep{daniels2008missing}. In principle, the choice of the assumptions to explore should be justified in light of the available information. However, in all reviewed studies, no reasonable justification was provided to support the choice of the alternative methods used in sensitivity analysis (often using CCA despite recognising its strong limitations). This is reflected by the relatively small number of studies providing full information about the missing data problem at hand, with the majority of the studies providing an average quality of missingness information (Figure~\ref{fig4}). Analysts may be able to improve current methodology through the adoption of more formal missing data strategies by taking into account the complexities of CEA data as well as a range of missing data assumptions. For example, MAR could be set as the benchmark assumption and external information may be incorporated into the model to elicit a set of MNAR departures from it~\citep{mason2018bayesian, leurent2018sensitivity, gabrio2019full}.

\subsection{Conclusions}\label{conclusions}
Given the complexity of the economic evaluation framework, the implementation of simple but likely inadequate analytic approaches may lead to imprecise cost-effectiveness results. This is a potentially serious issue for bodies such as ZIN in the Netherlands who use these evaluations in their decision making, thus possibly leading to incorrect policy decisions about the cost-effectiveness of new healthcare interventions.

Our review shows, over time, a change in many of the analysis components among standard practice in accordance with the recent ZIN's 2016 guidelines. This is an encouraging movement towards the standardised use of more suitable and robust analytic methods in terms of both statistical, uncertainty and missing data analysis. Nevertheless, improvements are still needed, particularly in the use of statistical software to implement advanced techniques as well as in the use of alternative missing data methods to explore plausible assumptions in sensitivity analysis.

\section*{Appendix}

\subsection*{Sample size distribution of reviewed studies}\label{A}

\begin{center}
FIGURE 5
\end{center}\vspace{0.5cm}

\subsection*{Missing data rates of reviewed studies}\label{B}

\begin{center}
FIGURE 6
\end{center}\vspace{0.5cm}

\bibliographystyle{apa} 
\bibliography{review_HTA_NL_bib}      

\newpage

\begin{table}[!h]
\centering
\scalebox{0.9}{
\begin{tabular}{cc|c|c}
  \toprule
 \multicolumn{2}{c|}{\textbf{Component}} & \textbf{2010-2015} ($n=96$) & \textbf{2016-2020} ($n=94$) \\ 
    & & $n (\%)$  & $n (\%)$ \\ 
\midrule
  \multirow{4}{*}{Perspective} & \textbf{societal} & 49 (51\%) & 41(44\%)  \\ 
  & healthcare/third party  & 24 (26\%) & 15 (16\%) \\ 
  & societal \& healthcare & 21 (23\%) & 37 (40\%) \\ 
  & unclear & 2 (2\%) & 0 \\
  \midrule
  \multirow{3}{*}{Analysis} & \textbf{CUA} & 30 (31\%) & 41 (44\%)   \\ 
  & CEA & 29 (30\%) & 16 (17\%) \\ 
  & CUA \& CEA & 37 (39\%) & 37 (39\%) \\ 
  \midrule
  \multirow{2}{*}{Design} & empirical & 86 (90\%) & 80 (83\%)  \\ 
  & model-based & 10 (10\%) & 14 (17\%) \\ 
  \midrule
 Horizon & $<1$ year$^\dagger$ & 25 (30\%) & 23 (29\%)  \\ 
 & $1$ year$^\dagger$  & 46 (53\%) & 41 (51\%) \\ 
  & $>1$ year$^\dagger$  & 16 (17\%) & 15 (20\%) \\ [0.5em]
  & \textbf{lifetime}$^\star$ & 7 (70\%) & 6 (43\%)   \\ 
  \midrule
  \multirow{1}{*}{Discounting} & relevant & 26 (27\%) & 25 (27\%)   \\ 
  (horizon $>1$ year) & \textbf{$4\%$ costs \& $1.5\%$ effects} & 15 (58\%) & 17 (68\%)  \\ 
  \midrule
  \multirow{1}{*}{Costs} &  \textbf{societal} & 51 (53\%) & 64 (68\%) \\ 
\midrule
  \multirow{1}{*}{Productivity losses} & \textbf{friction} & 20 (39\%) & 38 (59\%)   \\ 
    \multirow{3}{*}{(societal)} & human capital & 5 (10\%) & 6 (9\%) \\ 
  &friction \& human capital & 7 (14\%)  & 8 (13\%) \\ 
  & unclear & 14 (37\%) & 20 (19\%) \\ 
  \midrule
  \multirow{1}{*}{Quality of life} & \textbf{EQ-5D-5L} & 3 (4\%) & 9 (12\%)   \\ 
  \multirow{3}{*}{(CUA)} &  EQ-5D-3L & 16 (24\%) & 26 (33\%) \\ 
  & EQ-5D (unclear version) & 34 (51\%) & 22 (28\%) \\ 
  & other & 14 (21\%) & 21 (27\%) \\ 
  \midrule
  \multirow{4}{*}{Uncertainty analysis} & \textbf{CEP \& CEAC} & 60 (63\%) & 62 (66\%) \\ 
  & CEP & 24 (25\%) & 14 (15\%) \\ 
  & CEAC & 9 (9\%) & 14 (15\%) \\ 
  & none & 3 (3\%) & 4 (4\%) \\ [1em]
  & tornado$^\star$ & 3 (30\%) & 7 (50\%) \\ 
  \midrule
 Value of Information & \textbf{EVPI}$^\star$ & 1 (10\%)  & 1 (7\%) \\ 
  \bottomrule
\end{tabular}
}
\caption{Descriptive information of the reviewed studies for the periods 2010-2015 and 2016-2020. For each component of the economic evaluation, the approaches implemented are summarised and compared with the recommended approach from the 2016 guidelines (highlighted in bold). The superscripts $\dagger$ and $\star$ denote components for which proportions are calculated out of the total number of empirical and model-based analyses, respectively.}\label{tab1}
\end{table}

\begin{figure}[H]
\centering
\scalebox{0.6}{
\includegraphics{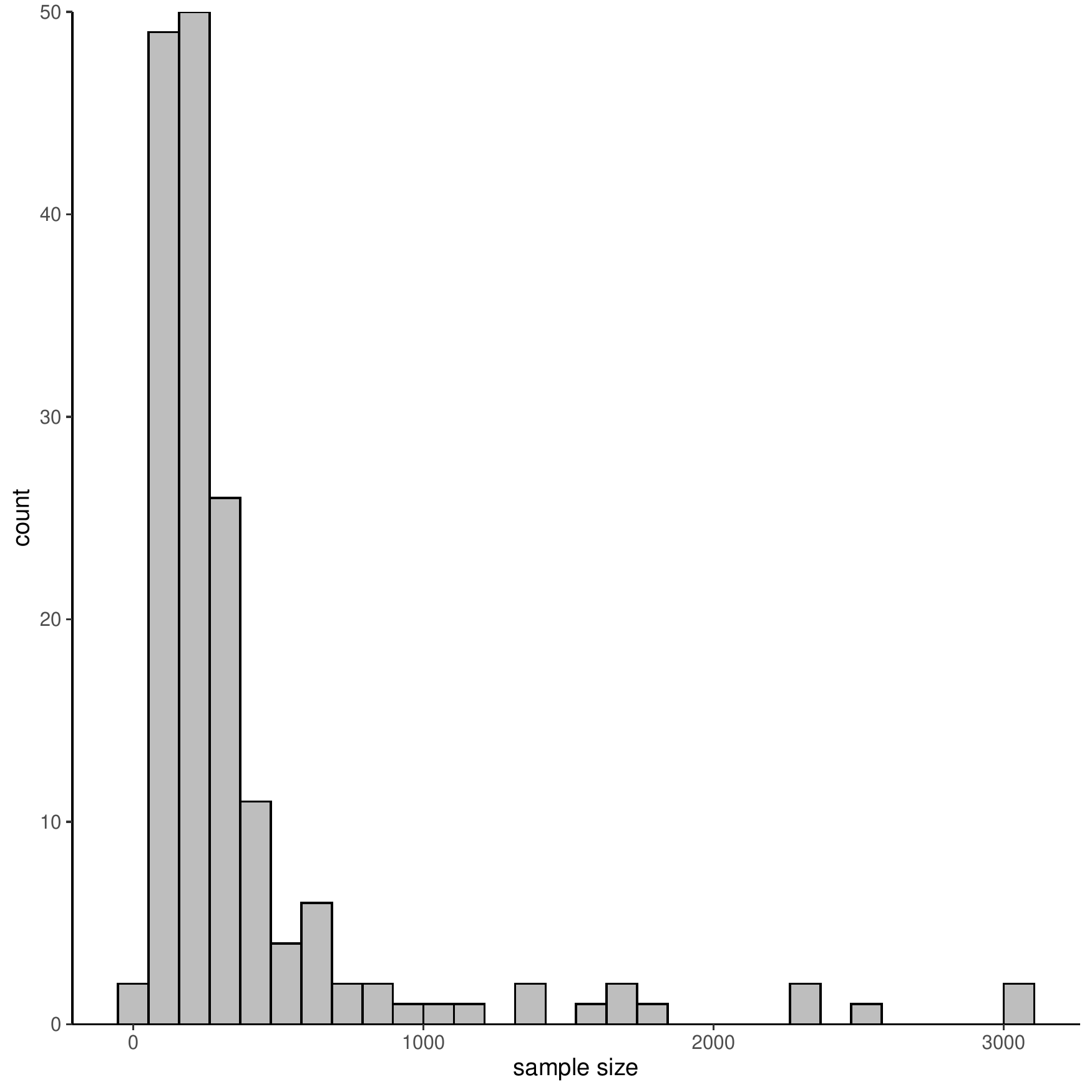}
}
\caption{Histogram of the sample size distribution of the $166$ empirical studies in included in the review.}\label{figA1}
\end{figure}

\begin{figure}[H]
\centering
\scalebox{0.6}{
\includegraphics{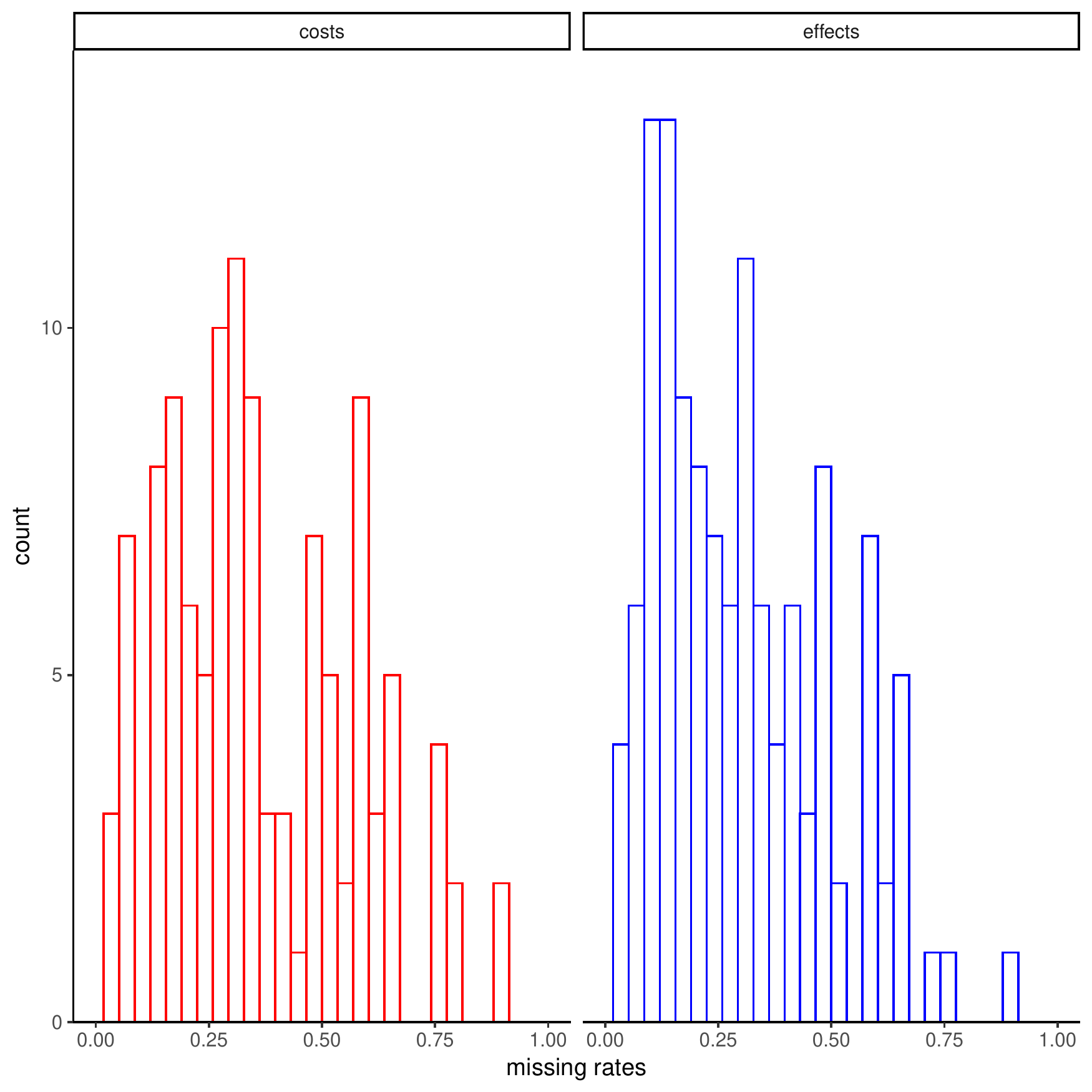}
}
\caption{Histograms of the distributions of missingness rates for effects (blue bars) and costs (red bars) among the empirical studies which provided the information.}\label{figA2}
\end{figure}


\end{document}